\documentclass[aps,prl,reprint,amsfonts,amsmath,superscriptaddress,floatfix,nofootinbib]{revtex4-1}
\usepackage{graphicx,epsfig,color,amsmath,booktabs}
\usepackage{soul}
\usepackage{mathtools}

\newcommand{\beqy}{\begin{eqnarray}}
\newcommand{\eeqy}{\end{eqnarray}}
\newcommand{\MSol}{{\rm M}_{\odot}}
\newcommand{\Tbase}{T_{\rm b}}
\newcommand{\Tcore}{T_{\rm core}}
\newcommand{\Qimp}{Q_{\rm imp}}
\newcommand{\Vn}{\mathbb{V}_n}
\newcommand{\VLn}{\mathbb{V}_{Ln}}
\newcommand{\Vcn}{\mathbb{V}^{(0)}_{cn}}
\newcommand{\Singlet}{$^1S_0$ }

\begin{document}

\title{Gapless neutron superfluidity can explain the late time cooling of transiently accreting neutron stars}
 
  \author{V. Allard}
 \affiliation{Institute of Astronomy and Astrophysics, Universit\'e Libre de Bruxelles, CP 226, Boulevard du Triomphe, B-1050 Brussels, Belgium}

 \author{N. Chamel}
 \affiliation{Institute of Astronomy and Astrophysics, Universit\'e Libre de Bruxelles, CP 226, Boulevard du Triomphe, B-1050 Brussels, Belgium}

\date{\today} 

\begin{abstract}
The current interpretation of the observed late time cooling of transiently accreting neutron stars 
in low-mass X-ray binaries during quiescence requires the suppression of neutron superfluidity in their crust 
at variance with recent ab initio many-body calculations of dense matter. 
Focusing on the two emblematic sources KS~1731$-$260 and MXB~1659$-$29, we show that their thermal
evolution can be naturally explained by considering the existence of a neutron superflow driven by 
the pinning of quantized vortices. Under such circumstances, we find that the neutron superfluid can  
be in a gapless state in which the specific heat is dramatically increased compared to that in the classical 
BCS state assumed so far, thus delaying the thermal relaxation of the crust. We have performed
neutron-star cooling simulations taking into account gapless superfluidity and we have obtained excellent
fits to the data thus reconciling astrophysical observations with microscopic theories. The imprint of 
gapless superfluidity on other observable phenomena is briefly discussed.
\end{abstract}

\maketitle

\section{Introduction}

Although neutron stars (NS) are formed in the furnace of gravitational-core collapse supernova explosions with
initial temperatures as high as $\sim 10^{11}-10^{12}$~K, they cool down very rapidly by releasing neutrinos 
so that their temperature drops down to $\sim 10^9$~K within a few days~\cite{potekhin2015}. Their extremely dense interior is
expected to become cold enough for the occurrence of quantum phase transitions not observed in any other
celestial bodies. Similarly to electrons in conventional terrestrial superconductors, free neutrons in the crust and possibly
in the outer core of a neutron star form \Singlet Cooper pairs, which condense at temperatures below 
$\sim 10^{10}$~K.  Predicted before the discovery of pulsars and only two years after the publication of 
the Bardeen-Cooper-Schrieffer (BCS) theory~\cite{migdal1959},  neutron superfluidity has since been corroborated by radio-timing 
observations of sudden spin-ups so called (frequency) glitches in numerous pulsars~\cite{antonopoulou2022}, interpreted as the
manifestation of the catastrophic unpinning of neutron quantized vortices~\cite{anderson1975,alpar1985}. 

However, superfluidity in the crust has been recently challenged by observations of NSs in low-mass X ray
binaries.  In these systems,   matter is transferred from a low-mass stellar companion to a NS via an
accretion disk.  The hydrogen-rich material that accumulates on the surface of the NS burns steadily
producing a thick helium layer.  Once the critical conditions for helium ignition are reached,  the overlying
envelope is converted into heavier nuclides within seconds. These thermonuclear explosions are observed as X-ray 
bursts lasting a few tens of seconds and with a recurrence time of 
hours to days~\cite{galloway2020}.  Less frequent but more energetic are superbursts lasting for a few hours 
with recurrence times of several years~\cite{Kuulkers2004},  presumably triggered by the unstable carbon burning~\cite{CummingBildsten2001,StrohmayerBrown2002}.

In most X-ray binaries,  accretion is not persistent but occurs sporadically~\cite{Bahramian2022}.  In particular, 
soft X-ray transients (SXTs) exhibit active periods of weeks to months separated by quiescent periods of 
years to decades.  So-called ‘quasipersistent’ SXTs remain active for years to decades.  As matter accumulates
on the NS surface,  ashes of X-ray bursts are buried and further processed due to electron captures, 
neutron captures and emissions, and possibly pycnonuclear fusions~\cite{meisel2018} releasing some heat in different parts of the crust.
In quasipersistent SXTs,  the
accretion can last long enough for the crust to be driven out of thermal equilibrium with the core.
Over the past two decades,  the thermal relaxation of a dozen SXTs has been monitored long enough after their 
outbursts ($\sim 10^3-10^4$ days) to probe all regions of the  crust~\cite{Wijnands2017}. The interpretation of the 
observed decline in temperature during the first few weeks and months requires some additional heating in the shallow 
layers of the crust~\cite{BrownCumming2009}  (see, e.g., Ref.~\cite{ChamelFantina2020} for a compilation of the inferred 
heat and references to proposed sources). 
The cooling at later times is dictated by the physics of the inner crust and neutron superfluidity~\cite{Page2012,Chaikin2018}. 
Observations of some SXTs, especially KS~1731$-$260 and MXB~1659$-$29, can hardly be explained by the standard cooling theory.

KS~1731$-$260 entered into a quiescent phase in 2001 after having accreted for 12.5 years. Its observation~\cite{Wijnands2001} 
provided the first direct evidence of the thermal relaxation of the NS crust~\cite{rutledge2002}. Monitoring campaigns 
of this source with Chandra and XMM-Newton satellites  confirmed this scenario~\cite{wijnands2002,cackett2006,shternin2007,BrownCumming2009}. 
However, later observations~\cite{Cackett2010} revealed that this source had become colder than expected. 
MXB~1659$-$29 was monitored in quiescence after an accretion outburst of 2.5 years~\cite{Wijnands2003,wijnands2004,cackett2006,cackett2008}. 
The data were modelled in Ref.~\cite{BrownCumming2009}. Observations taken 11 years after outburst~\cite{Cackett2013} showed an unexpected drop of luminosity. This could be explained by an increased hydrogen column density $N_H$ on the line of sight due to precession of the accretion disk~\cite{Cackett2013}. Alternatively, these observations suggested that the thermal equilibrium between the crust and the core had not been restored. Based on classical molecular dynamics simulations, it was proposed that the densest layers 
of the crust have a low thermal conductivity~\cite{horowitz2015}. But quantum molecular dynamics simulations performed later did not support this possibility~\cite{nandi2018}. The data of both sources were also analyzed in Ref.~\cite{Turlione2015} and the best fits were achieved 
by artificially suppressing superfluidity in most part of the crust.
In 2015, MXB~1659$-$29 went back into outburst~\cite{sanchez2015}, which lasted 1.7~years~\cite{Parikh2019}. No significant variations of $N_H$ that would confirm an hypothetical precession of the accretion disk were found. In 2016, Merritt et al.~\cite{Merritt2016} reported observations of KS~1731$-$260 14.5 years after outburst and were able to fit the data (with a rather large $\chi^2$) using the small neutron pairing gaps of Ref.~\cite{Schwenk2003}. Deibel et al.~\cite{Deibel2017} obtained very good fits for both KS~1731$-$260 and MXB~1659$-$29 considering that neutrons remain normal in the deep crust and in the outer core, based on extrapolations of quantum Monte Carlo calculations~\cite{Gandolfi2008} (MXB~1659$-$29 was further studied in Refs.~\cite{Parikh2019,potekhin2021,lu2022,potekhin2023} but the observations reported in Ref.~\cite{Cackett2013} were discarded). 
However, more recent calculations have ruled out this possibility~\cite{Gandolfi2022}, and results are now 
consistent with those from other approaches~\cite{cao2006,drissi2022,Krotscheck2023} (see \cite{supplement}). Besides, superfluidity in both the crust and the outer core is independently required for the interpretation of pulsar
glitches~\cite{andersson2012,chamel2013,ho2015,pizzochero2017}. Such phenomena have been detected in accreting NSs as well~\cite{Galloway2004,serim2017}. 

In this letter,  we show how those apparently contradictory observations can be reconciled by considering the existence of a superflow 
in accreted NS crusts. In particular, we contemplate the possibility that the superfluid is gapless: the energy spectrum of 
quasiparticle excitations is continuous 
whereas the (complex) order parameter (whose modulus coincides with the pairing gap in the absence of superflow) remains finite.  The microscopic 
theory is presented in Ref.~\cite{AllardChamel2023PartI}. The absence of a gap translates into a neutron specific heat that is orders of magnitude 
larger than that predicted by the classical BCS theory.  
The impact of gapless superfluidity on the late time cooling of SXTs is studied, focusing on the emblematic sources MXB~1659$-$29 and KS~1731$-$260.  After briefly describing our model, NS cooling simulations are presented and discussed. Finally, we mention other observational phenomena that could confirm the existence of gapless superfluidity in NSs.

\section{NS cooling model}
\label{part:SpecificHeatGapless}

The thermal evolution of SXTs is followed using the code \texttt{crustcool}\footnote{\url{https://github.com/andrewcumming/crustcool}}, which solves the heat diffusion equation in the NS crust assuming a constant gravity~\cite{BrownCumming2009}. This code, based on the accreted-crust model of Ref.~\cite{HaenselZdunik1990}, was previously employed in Refs.~\cite{Cackett2013,horowitz2015,Deibel2017} to analyze the same sources. Shallow heating is accounted for by adjusting the temperature $\Tbase$ at the bottom of the envelope at the column depth of $10^{12}$~g~cm$^{-2}$  (see also Refs.~\cite{Turlione2015,Deibel2015,cumming2017}). 

Gusakov and Chugunov~\cite{GusakovChugunov2020,GusakovChugunov2021} have recently shown that the diffusion of 
superfluid neutrons in accreting NS 
crusts changes the composition and the equation of state. Moreover, the nuclear heating is substantially
reduced. We have modified the \texttt{crustcool} code accordingly (see \cite{supplement}). More importantly, 
we have implemented more realistic microscopic neutron pairing calculations and allowed for gapless superfluidity. In the normal phase at temperatures $T$ much lower than the neutron Fermi temperature, the neutron 
specific heat is approximately given by (with $k_B$ Boltzmann's constant and $\hbar$ the Planck-Dirac constant)
\beqy\label{eq:CV-normal}
c_{N}^{(n)}(T)\approx \frac{1}{3}\frac{k_{Fn}m_n^\oplus}{\hbar^2} k_{\text{B}}^2 T \, ,
\eeqy
where $m_n^\oplus$ is the neutron effective mass, which can be approximated by the bare neutron mass $m_n$~\cite{cao2006}, and $k_{Fn}$ is the neutron Fermi wave number. 
In the superfluid phase and in the absence of superflow,  which we will refer to as the classical BCS state, the neutron specific heat is exponentially suppressed 
\beqy\label{eq:CV-BCS}
c_{S}^{(n)}(T<T^{(0)}_{cn}) = R_{00}^{\rm (BCS)}(T/T^{(0)}_{cn}) c_{N}^{(n)}(T)  \, .
\eeqy
The factor $R_{00}^{\rm (BCS)}(T/T^{(0)}_{cn})$ is given in Ref.~\cite{YakovlevandLevenfish1994}.
The critical temperature $T_{cn}^{(0)}$ is determined by the order parameter $\Delta_n^{(0)}$ at $T=0$ through 
the BCS relation $k_{\rm B}T_{cn}^{(0)}=\exp(\gamma)\Delta_n^{(0)}/\pi\simeq 0.56693\Delta_n^{(0)}$ ($\gamma\simeq 0.57722$ being the Euler-Mascheroni constant). 
In presence of superflow, the order parameter becomes complex and no longer represents the gap in the quasiparticle energy spectrum~\cite{AllardChamel2023PartI}. The gradient of its phase $\phi_n$ defines the superfluid velocity as $\pmb{V_n}=\hbar/(2m_n) \pmb{\nabla}\phi_n$. 
The effects of the superflow on $c_{S}^{(n)}$ are governed by some effective neutron superfluid velocity $\Vn$. 
At densities prevailing in the crust of NSs, $\Vn \approx V_n$~\cite{ChamelAllard2021}.  
For $\Vn<\VLn\approx \Delta_n^{(0)}/(\hbar k_{Fn})$, no quasiparticles are present and $c_{S}^{(n)}$ remains exponentially suppressed as in the BCS limit corresponding to $\Vn=0$.  
Expressions can be found in Ref.~\cite{AllardChamel2023PartI}.  
For $\VLn\leq \Vn\leq \Vcn\approx \exp(1) \VLn/2$,  the neutron superfluid is gapless (the modulus $\Delta_n$ of the order parameter remaining finite) and $c_{S}^{(n)}$ is only moderately reduced compared to that in the normal phase. At $T\ll T_{cn}^{(0)}$, the reduction factor is essentially independent of the temperature and is given by~\cite{AllardChamel2023PartI} 
\beqy\label{eq:Gapless-Suppression}
R_{00}^{\rm (Gapless)}(\Vn)=\sqrt{1-\left(\frac{2}{\exp(1)}\frac{\Delta_n}{\Delta_{n}^{(0)}}\frac{\Vcn}{\Vn}\right)^2}\, ,
\eeqy
with~\cite{ChamelAllard2021}
\beqy
\Delta_n (\Vn) &=& 0.5081\Delta_{n}^{(0)}\sqrt{1-\frac{\Vn}{\Vcn}} \notag \\ 
&\times& \left(3.312\frac{\Vn}{\Vcn}-3.811\sqrt{\frac{ \Vcn}{\Vn}} +5.842\right)\, .
\eeqy
In the gapless state, the neutron specific heat is a \emph{universal} function of $\Vn/\Vcn$ or equivalently of $\Vn/\VLn$, i.e. it is independent of the adopted results for $\Delta_n^{(0)}$. 
Neutron superfluidity is destroyed (i.e.  $\Delta_n=0$) when $\Vn\geq \Vcn$ and the neutron specific heat then reduces to Eq.~\eqref{eq:CV-normal}.  The actual value of $\Vn$ depends on the dynamical evolution of the specific SXT under consideration.  In the following, we will treat $\Vn/\VLn$ as a free parameter.

\section{Results and discussions}
\label{part:NumericalResults}

We now discuss the cooling of KS~1731$-$260 and MXB~1659$-$29. 
Observational data and full numerical results are given in \cite{supplement}. 
As in previous studies~\cite{BrownCumming2009,horowitz2015,Merritt2016,Deibel2017,brown2018}, we assume in both cases a constant accretion rate $10^{17}$~g~s$^{-1}$ consistent with the time-averaged accretion rate found in Ref.~\cite{Galloway2008}.  A variable accretion rate can affect the cooling,  but not in the late stage of interest here~\cite{ootes2016}.  We set the NS mass to $1.62\MSol$ and the radius to $11.2$~km, as in Refs.~\cite{BrownCumming2009,horowitz2015,brown2018,Deibel2017}. Unless stated otherwise, we adopt the realistic pairing calculations of Ref.~\cite{Gandolfi2022}. To estimate the uncertainties in the parameters $\Tbase$, $\Tcore$, $\Qimp$ and $\Vn$, we have run Markov Chain Monte Carlo simulations~\cite{supplement}.

\begin{figure}[h]
\includegraphics[width=8.5cm]{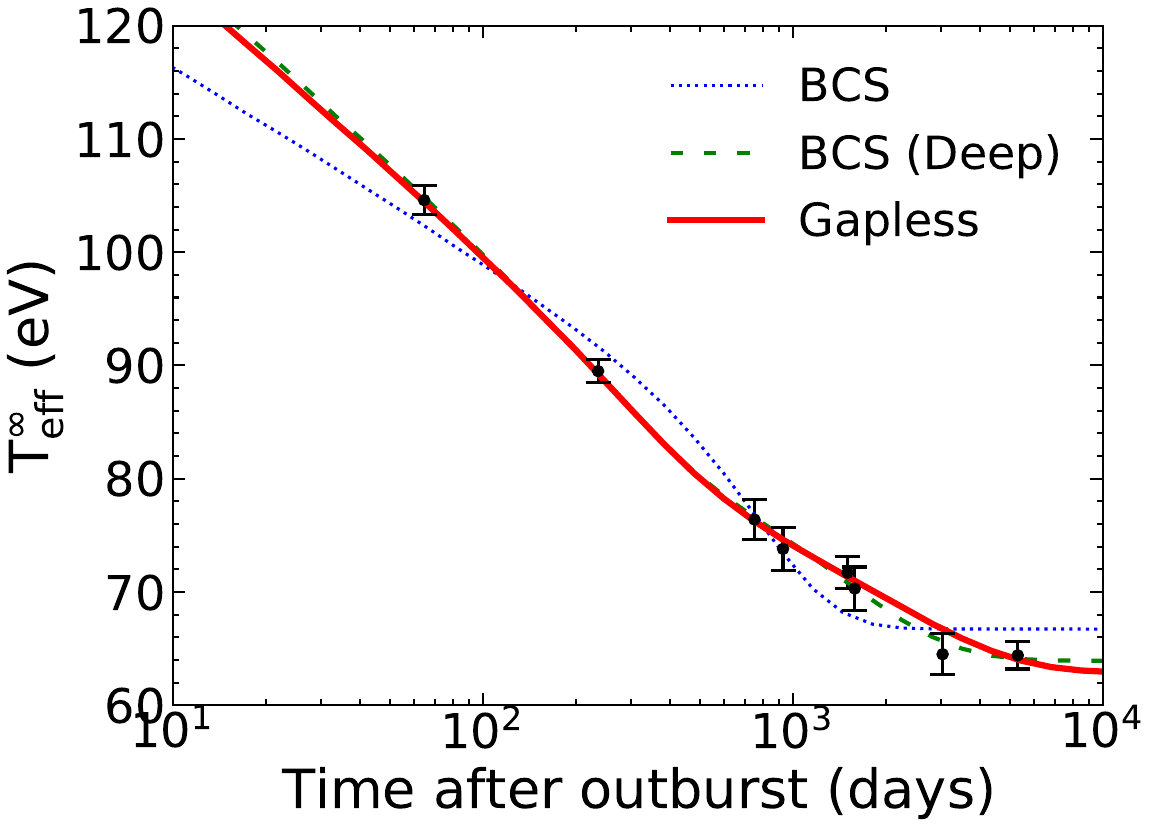}
\caption{Evolution of the effective surface temperature in electronvolts of KS~1731$-$260 (as seen by an observer at infinity) as a function of the time in days after 12.5 year outburst. Symbols represent observational data with error bars. 
The dotted and solid lines are models considering BCS and gapless superfluidity respectively using the realistic pairing calculations of Ref.~\cite{Gandolfi2022}. The dashed line was obtained assuming BCS superfluidity with the fine-tuned ``Deep'' gap of Ref.~\cite{Turlione2015}. 
}
\label{fig:KSG22-Opti}
\end{figure}

Results for the thermal evolution of KS~1731$-$260 after the 12.5 years of outburst are displayed in 
Fig.~\ref{fig:KSG22-Opti}. Ignoring the presence of superflow as in previous studies, this model 
(dotted curve) fails to explain the late time cooling after $10^3$ days and leads to a rather 
poor fit of the earlier observations. The optimum parameters with uncertainties at 68\% level are
$\Qimp=10.56^{+2.15}_{-1.97}$, $\Tbase=2.45^{+0.19}_{-0.18}\times 10^8$~K 
and $\Tcore=(4.69\pm0.14)\times 10^7$~K. 
The last four data points can only be reproduced by artificially fine tuning the pairing gap (dashed curve), 
as in Ref.~\cite{Turlione2015}. 

In contrast, allowing for gapless superfluidity (solid curve) yields an excellent fit to the full data 
set. Our best model is obtained for $\Vn=1.21^{+0.10}_{-0.11}\VLn$. However, the distribution of $\Vn$ 
is rather broad and the 95\% credibility interval extends down to about $\VLn$. The other parameters are 
$\Qimp=5.80^{+1.68}_{-1.25}$, $\Tbase=3.13^{+0.19}_{-0.20}\times 10^8$~K and $\Tcore=3.99^{+0.26}_{-0.34}\times 10^7$~K. 

\begin{figure}[h]
\includegraphics[width=8.5cm]{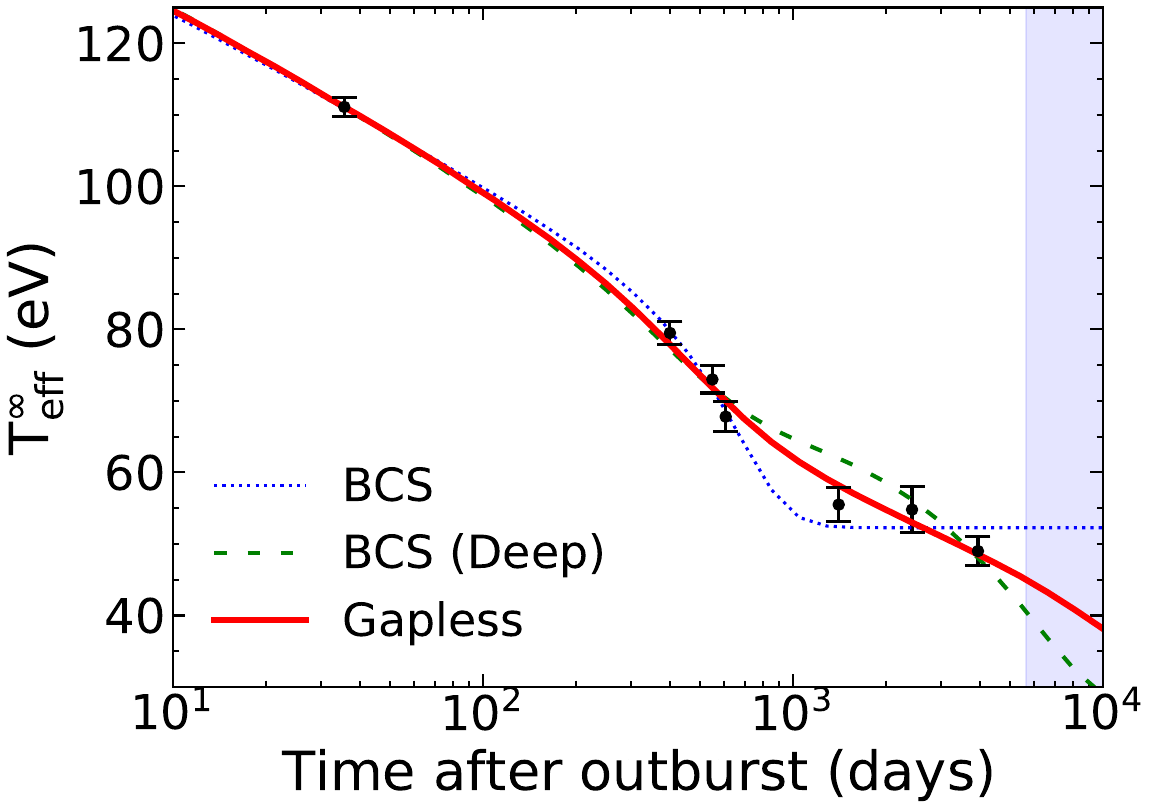}
\caption{Same as Fig.~\ref{fig:KSG22-Opti} for MXB~1659$-$29 after the first outburst. The shaded area corresponds to the second accretion phase (which occurred in 2015) and its subsequent cooling phase: the cooling curves within this region depict the expected behavior had this outburst not occurred.}
\label{fig:MXB22-Opti}
\end{figure}

Figure~\ref{fig:MXB22-Opti} shows results for the first outburst of MXB~1659$-$29. Restricting to 
BCS superfluidity with realistic pairing, 
the best model (dotted curve) is obtained for $\Qimp=8.05^{+0.71}_{-0.61}$, $\Tbase=(3.11\pm 0.13)\times10^8$~K,  and  
$\Tcore=(2.69\pm 0.15)\times10^7$~K. 
With this model, the NS is cooling so rapidly that thermal equilibrium is restored about $10^3$ days after the end of the outburst thus failing to reproduce the last data point. This puzzle can be naturally solved by taking into account the superflow. Our best model  (solid curve) is found for $\Vn=1.23^{+0.09}_{-0.11}\VLn$, $\Qimp=17.16^{+3.49}_{-3.70}$, $\Tbase=3.14^{+0.16}_{-0.15}\times 10^8$~K and $\Tcore=0.85^{+0.59}_{-0.54}\times 10^7$~K. 
To check the consistency of our model, we have analyzed the second outburst keeping fixed the core temperature. 
As discussed in Ref.~\cite{Parikh2019}, variations of $\Tcore$ between the two outbursts and during the subsequent 
crust cooling are expected to lie within the observational uncertainties. Our model reproduces the observations very well~\cite{supplement}. We find no significant change of $\Qimp$ (consistent with the analysis of Ref.~\cite{Parikh2019} in the standard paradigm) and $\Vn$ at the 95\% level, contrary to $\Tbase$. However, $\Tbase$ (related to shallow heating) needs not remain the same 
between outbursts. For completeness, we have also shown the best-fit model with the ``Deep'' gap of Ref.~\cite{Turlione2015}. However this gap, which was empirically adjusted to fit the cooling data of KS~1731$-$260
within the traditional model of accreted NSs of Haensel\& Zdunik~\cite{HaenselZdunik1990}, does not  
provide satisfactory results for the first outburst of MXB~1659$-$29.

For both sources, by allowing for gapless superfluidity we have obtained excellent fits to the data without having to introduce a highly disordered layer in the deep crust in agreement with quantum molecular dynamics simulations~\cite{nandi2018}. Running simulations within the traditional model of accreted NSs~\cite{HaenselZdunik1990}, we have found that the diffusion of superfluid neutrons introduced in Refs.~\cite{GusakovChugunov2020,GusakovChugunov2021} does not solve in itself the puzzle of the late time cooling~\cite{supplement}.
At the time of this writing, KS~1731$-$260 and MXB~1659$-$29 are still in quiescence~\cite{Maccarone2022}. According to our best models, the crust of the former has finally reached  thermal equilibrium whereas the crust of the latter is further cooling.

The presence of a finite superflow in NS crusts, as suggested by our best cooling models of MXB~1659$-$29 and KS~1731$-$260, is not unexpected. 
During accretion episodes, the crust and all particles strongly coupled to it (constituting most of the star) are expected to be spun up due to the transfer of angular momentum from the in-falling material. 
This so called 'recycling' scenario proposed to explain the existence of millisecond pulsars~\cite{alpar1982,radhakrishnan1982} has been recently confirmed by the discovery of accreting millisecond X-ray pulsars~\cite{patruno2021,Salvo2022} and transitional millisecond pulsars~\cite{papitto2022}. Evidence for the fact that both KS~1731$-$260 and MXB~1659$-$29 have been recycled during their history come from observations of X-ray burst oscillations at high frequency, respectively $\sim$524~Hz and 567~Hz~\cite{Smith1997,Wijnands2001,Wijnands2003,Galloway2008}, likely related to the NS spin frequency. Due to pinning of quantized vortices, the neutron superfluid velocity is locked\footnote{The quantization of the neutron superflow implies that $\int \pmb{V^\prime_n}\cdot \pmb{{\rm d}\ell}=N h/(2m_n)$ along any contour enclosing $N$ vortices. Here $\pmb{V^\prime_n}=\pmb{V_n}+\pmb{v_N}$ is the neutron superfluid velocity in a fixed external frame in which the star is rotating with the velocity $\pmb{v_N}$. If vortices are pinned, $\pmb{V^\prime_n}$ must therefore remain unchanged.} so that in the crust frame $V_n$ therefore also $\Vn$ both increase. 
At the end of an outburst and during the quiescent period that follows, $\Vn$ is likely to remain essentially constant unless vortices unpin; unlike isolated pulsars, NSs in low-mass X-ray binaries have typically very weak  magnetic fields $\sim 10^8-10^9$~G so that the spin down caused by electromagnetic braking is not expected to be very effective~\cite{Burderi2002}. This justifies our assumption of a constant superflow throughout the thermal relaxation.

\section{Conclusions}

Gapless neutron superfluidity in the inner crust of NSs provides a natural explanation for the 
observed late time cooling of quasipersistent SXTs due to the huge enhancement of the neutron specific heat compared to that in the classical BCS case. The neutron specific heat can thus be comparable to that in the normal phase without requiring the unrealistic suppression of superfluidity as previously proposed. 
Focusing on the emblematic sources KS~1731$-$260 and MXB~1659$-$29 for which the interpretation via the standard cooling theory has been challenged, we have obtained excellent fits to the observational data 
using realistic neutron pairing calculations~\cite{Gandolfi2022} and without introducing a highly disordered layer at the crust bottom in agreement with quantum Monte-Carlo calculations~\cite{nandi2018}. We have also checked the consistency of our model between the two outbursts of MXB~1659$-$29. According to our simulations, the crust of KS 1731$-$260 is now in thermal equilibrium with the core, whereas MXB~1659$-$29 is still cooling (at variance with the standard paradigm~\cite{supplement}). These predictions could be tested by future observations and could provide more stringent constraints on $\Vn$.

Gapless superfluidity is driven by the presence of a superflow in the crust, as expected to arise from the pinning of quantized vortices. 
The lag between the superfluid and the rest of star is limited by the strength of pinning forces acting on individual vortices. The maximum superfluid velocity can be estimated as $V_{\rm cr}\sim 10^7(f_p/10^{18}~{\rm dyn/cm})$~cm/s \cite{pizzochero2011}, 
where $f_p$ denotes the average (on the appropriate hydrodynamic scale) pinning force per unit length. Systematic fully microscopic calculations of the force on a single pinning site remain computationally challenging~\cite{wlazlowski2016}. Averaging over many pinning sites could lead to much stronger forces~\cite{Link2022}. Current estimates for $f_p$ are at most of order $10^{18}$ dyn/cm leading to $V_{\rm cr}\sim 10^7$ cm/s; $\VLn$ is found to be somehow higher, of the order of $10^8$~cm/s. However, experiments using cold atoms~\cite{miller2007} suggest that Landau's velocity could be significantly suppressed by the presence of clusters, which we have ignored here. 
Moreover, vortices extend to the core where they can pin to proton fluxoids thus further increasing $f_p$ hence also $V_{\rm cr}$. It is therefore not inconceivable that $\VLn \lesssim V_{\rm cr}$. The excellent fit of the cooling data from SXTs brings support to this hypothesis and calls for further studies of the 
vortex dynamics in NSs.

The existence of a superflow in the crust could potentially have other observational consequences. 
At some point most likely during outburst, vortices may be unpinned (e.g. due to thermal activation~\cite{linkepstein1996}) leading to a sudden spin up of the superfluid accompanied by a spin down of the star~\cite{ducci2015}.  This will be manifested by an anti-glitch, i.e. a decrease of the spin frequency,  or possibly a glitch under certain circumstances~\cite{Kantor2014}. 
Whether such an event occurred in MXB~1659$-$29 is difficult to assess due to the comparatively large uncertainties in the spin frequency measured from X-ray burst oscillations. The same difficulty will arise when KS~1731$-$260 will return to outburst. These sources are nuclear-powered X-ray pulsars exhibiting X-ray bursts due to thermonuclear explosions.  More accurate measurements of the spin frequency are made in accretion-powered X-ray pulsars undergoing channeled accretion due to magnetic fields. 
Among them Aql X-1,
whose accurately measured spin frequency is 550.2744 Hz~\cite{casella2008}, has a prolific activity with 23 outbursts from 1996 to 2015 and the thermal emission during quiescence has been observed~\cite{Ootes2018}.
However, the periods between outbursts only last for weeks and are too short to probe the deep crust and superfluidity. 
The analysis of the observational data is further complicated by low-level residual accretion during quiescence~\cite{Waterhouse2016}. A more promising source to test our scenario is HETE~J1900.1$-$2455 with a spin frequency of 377.296171971(5) Hz~\cite{Kaaret2006}. This pulsar has recently been observed after the end of a 10-year long accretion outburst and appears unusually cold~\cite{Degenaar2017,degenaar2021}. Assuming that the crust has fully relaxed, the standard cooling theory requires the suppression of nucleon superfluidity in the core~\cite{degenaar2021} at variance with theoretical expectations. The analysis of this source is left for future studies.

\begin{acknowledgments}
This work was financially  supported by the Fonds de la Recherche Scientifique (Belgium) under Grants 
No. PDR T.004320 and IISN 4.4502.19. We thank Prof A. Sedrakian, N. Shchechilin and L. Planquart for discussions. 
\end{acknowledgments}

\bibliography{references.bib}

\end{document}